# Frequency-domain simulations of a negative-index material with embedded gain

Yonatan Sivan,[1,2,*] Shumin Xiao,[1] Uday K. Chettiar,[1,3] Alexander V. Kildishev,[1] and Vladimir M. Shalaev[1]

[1]*Birck Nanotechnology Center, School of Electrical and Computer Engineering, Purdue University, West Lafayette, IN 47907, USA*
[2]*Department of Electrical Engineering, The State University of New York at Buffalo, Buffalo, NY 14260, USA*
[3]*Electrical and Systems Engineering, University of Pennsylvania, Philadelphia, PA 19104, PA*
*[*]ysivan@purdue.edu*

**Abstract:** We solve the equations governing light propagation in a negative-index material with embedded nonlinearly saturable gain material using a frequency-domain model. We show that available gain materials can lead to complete loss compensation only if they are located in the regions where the field enhancement is maximal. We study the increased enhancement of the fields in the gain composite as well as in the metal inclusions and show analytically that the effective gain is determined by the *average* near-field enhancement.

## References and links

## 1. Introduction

In recent years, the field of nanoplasmonics has experienced fast growth and has enjoyed a large share of the research in the areas of optics and solid-state physics. This huge interest originates from two basic aspects of plasmon excitation. First, being evanescent in nature, plasmonic excitations are not diffraction-limited; hence, they can give rise to light concentration over distances much smaller than the wavelength. This phenomenon enables improved imaging and sensing, improved light storage and nanolithography, optical device miniaturization and many more applications. Simultaneously, plasmonic resonances give rise to strong local field enhancements that can be up to several orders of magnitude. This phenomenon can lead to dramatic increase in efficiency of various optical phenomena, such as fluorescence, absorption surface-enhanced Raman scattering (SERS) spectroscopy, as well as nonlinear processes, such as high-harmonic generation or multi-photon absorption.

One of the most fascinating branches of nanoplasmonics is the emerging field of plasmonic metamaterials (see e.g., [1–3] for recent reviews). Metamaterials are man-made, metal-dielectric, subwavelength structures designed to have prescribed electromagnetic properties, especially properties that cannot be found in nature such as optical magnetism [4,5], a negative refractive index [6–10] and hyperbolic (indefinite) dispersion [11], to name a few. Among the fascinating applications of metamaterials are subwavelength imaging [12–14], invisibility cloaking [15,16], improved photovoltaics [17] and nano-lasers [18–23]. However, the performance of optical plasmonic devices is limited by the strong absorption losses from the metal inclusions, thus hindering their applicability in commercial devices. In particular, the realization of the exceptional electromagnetic properties associated with metamaterials requires a significant reduction of these losses.

Several methods have been proposed to reduce, avoid or overcome the losses in metamaterials, such as optical-parametric amplification [24,25], electromagnetically-induced transparency [26–28] and time-reversal by negatively refracting/reflecting nonlinear interfaces [29]. However, the most explored method so far to reduce the losses is to incorporate gain media into the metamaterial design. Indeed, partial and even complete compensation of absorption losses was predicted theoretically and demonstrated experimentally for basic plasmonic configurations such as surface plasmon-polaritons and plasmon-polariton waveguides [30–40], coated nano-particles [19,22,41] and nano-particle aggregates [42]. Further theoretical works discussed loss compensation and lasing by gain materials for various plasmonic/metamaterial structures, see e.g [20–23,43–47]. to name a few. Some works [20,46,47] showed that the *effective* gain in plasmonic devices can be much higher than the gain exhibited by the bulk gain material. This observation was attributed to the near-field enhancement inherent to the plasmonic response.

Many of the previous studies used a phenomenological gain mechanism by setting a constant gain coefficient. In [46,47], Fang *et al.* used a self-consistent *time-domain* model of a standard gain system composed of two coupled Lorentz oscillators to study loss compensation in several metamaterial designs. In the proposed configuration, the first oscillator is pumped, thus creating a population inversion that is then used to amplify a delayed probe pulse.

In this paper, we present an alternative, self-consistent, *frequency-domain* model of a general gain system [19,34–38] and show that it may be well-suited for the study of loss compensation in negative-index materials (NIMs) in a pump-probe configuration. As always, the frequency-domain formulation is much more computationally efficient compared with the time-domain model; moreover, it is not limited to Lorentzian lineshapes.

Generically, gain systems are described by a saturable absorptive response, i.e., they are inherently nonlinear. Accordingly, the gain induced by the pump light may not be uniform in space. We solve the nonlinear equations governing the *pump*-field distribution exactly, thus, establishing the exact spatial dependence of the gain experienced by the probe. To the best of our knowledge, these are the first such simulations in the context of plasmonics. We then solve the equations governing the probe-field distribution under the assumption that the probe-field is weak and does not saturate the gain. While under these conditions the probe equations are linear, the case of a strong saturating probe can be solved within our formalism in a similar manner to the solution of the nonlinear pump equations.

As in previous works [20,46,47], we observe that the loss compensation is more efficient than expected from an estimate based on the bulk parameters of the gain composite. Using exact relations derived from the Poynting theorem and rate equations, we show that the compensation is directly related to the *average* near-field enhancement. As a consequence, we show that complete compensation can be obtained for emitters with realistic gain coefficients if they are incorporated into the spacer layer where the field enhancement is maximal. Indeed, simulations of a standard negative-index fishnet design show that the gain compensation can be an order of magnitude more efficient for an emitter placed in the spacer layer compared with an emitter placed above or below the structure. Furthermore, we show that an estimate of the gain required for complete compensation should also take into account the increased losses in the metal due to the field enhancement in the metal.

## 2. Frequency-domain model - pump-probe configuration

Compensation of the strong absorption losses in plasmonic devices requires the most efficient emitters. Among these are semiconducting nanocrystals (quantum dots) [48], semiconducting polymers [49], rare-earth doped glasses [50] and dye molecules [51]. The latter are typically described as two broadened levels where the Einstein coefficients for stimulated absorption and emission are generalized through the corresponding cross sections, see Fig. 1(a). Frequently, the vibrational-level bands are replaced by a single higher level, thus constituting a four-level system (4LS), see Fig. 1(b).

The 4LS description is also appropriate to quantum dots and rare-earth doped glasses, and it is qualitatively and quantitatively similar to the broadened level description. Under this description, electrons are pumped from the ground state (level 0) to the excited band (or level 3), and they then quickly relax to the lowest level of the first excited band (or level 2). This level, henceforth denoted as the lasing level, is characterized by a long lifetime with respect to all other timescales in the system. This allows electrons to accumulate in the lasing level and consequently to create a population inversion between the higher and lower bands (or levels 2 and 1). This inversion can then be used to amplify a signal via stimulated emission.

In principle, the evolution of the level population should be described in the time-domain. This requires a self-consistent solution of the rate equations with the Maxwell equations in which the polarization has the standard, causal, Drude and/or Lorentz forms which are compatible with the 4LS model [52,53]. Nevertheless, for a sufficiently strong pump, the system approaches its steady-state population after a small fraction of the lasing level lifetime $\tau$. Indeed, under these conditions, the rate of photons emitted through stimulated emission, given by $\sigma_{em} I / \hbar \omega$, can be much higher than the rate of spontaneous emission, given by $1/\tau$ [52,53]. The population will then approach steady-state at an exponential rate given by $\sigma_{em} I / \hbar \omega \lessgtr 1/\tau$ with spontaneous emission causing small oscillations (residual transient solutions). This behaviour was explicitly demonstrated in the time-domain simulations given in e.g [47]. Thus, at an excellent approximation, a sufficiently short (and weak, see below) probe, sent after the steady-state is reached, will experience the constant steady-state population inversion. In this case, the dynamics of the probe can be accurately described by a frequency-domain model, which is far more efficient in terms of computation time compared with the time-domain model. Furthermore, a frequency-domain model can account for more general dispersion profiles, such as those of dye molecules which deviates strongly from a Lorentzian shape. Such frequency-domain models have been widely used for dye lasers [51] and more recently, in the context of surface plasmon amplification [34–38] and nano-particle lasers [19].

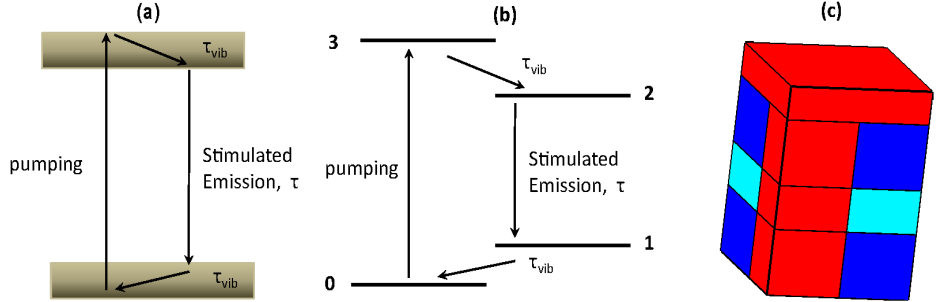

Fig. 1. (Color online) (a) Schematic of stimulated emission from a pumped, broadened two-level system. (b) Same as (a) for a four-level system. (c) Schematic illustration of a quarter unit cell of the fishnet structure studied in simulations. Silver components are shown in blue, spacer layer in light-blue and the gain-composite in red. In some simulations below (sample 3), the gain is also included in the spacer.

The populations of the various levels are determined from the steady-state solutions to the rate equations. If, indeed, the lifetime of the higher vibrational levels (or levels 3 and 1) $\tau_{vib}$ is very short compared to the lasing lifetime $\tau$ and the rate of stimulated emission from level 3, then the steady-state populations in the higher vibrational levels are essentially zero, and the populations of the lasing and ground-state levels are given by ($N_0 = N - N_2$)

$$N_2 = N \frac{L_{abs}(\lambda_{pump}) |\bar{E}_{pump}(\lambda_{pump})|^2}{1 + L_{abs}(\lambda_{pump}) |\bar{E}_{pump}(\lambda_{pump})|^2 + q_i L_{em}(\lambda) |\bar{E}(\lambda)|^2}, \qquad (1)$$

where N is the emitter concentration, the normalized lineshapes, $L_{abs}(\lambda)$ and $L_{em}(\lambda)$ are related to the absorption and emission cross sections through

$$\begin{aligned}
\sigma_{abs}(\lambda) &= L_{abs}(\lambda)\sigma_{abs,0}, & L_{abs}(\lambda_{30}) &= 1, \\
\sigma_{em}(\lambda) &= L_{em}(\lambda)\sigma_{em,0}, & L_{em}(\lambda_{21}) &= 1,
\end{aligned} \quad (2)$$

with $\lambda_{30}$ and $\lambda_{21}$ being the central wavelengths of absorption and emission, respectively; the cross-section ratio $q_i \approx \dfrac{\sigma_{em,0}}{\sigma_{abs,0}}$ is the quantum yield and $|\bar{E}_{pump}| = |E_{pump}/E_{sat}|$ and $|\bar{E}| = |E/E_{sat}|$ where $E_{pump}$ and $E$ denote the electric field amplitude of the pump and probe pulses (an exponential dependence of the form $e^{-i\omega t + i\vec{k}\cdot\vec{r}}$ is assumed), normalized with the saturation field magnitude $|E_{sat}|$, given by

$$|E_{sat}|^2 \equiv \frac{4h}{\epsilon_0 n' \lambda_{30}} \frac{1}{\tau \sigma_{abs,0}},$$

where $h$ is the Planck constant, $\epsilon_0$ is the permittivity of vacuum, and $n'$ is the refractive index of the gain-composite (consisting of the material hosting the emitters and the emitters themselves).

Equation (1) shows that as long as $q_i|E|^2 \ll |E_{pump}|^2$, the probe field has a negligible effect on the steady-state populations. This was explicitly demonstrated via time-domain simulations in [46,47], see also discussion in Section 5.

Assuming the host material is transparent, the absorption and gain coefficients, defined by

$$\begin{aligned}
\alpha(\lambda, \lambda_{pump}) &= \sigma_{abs}(\lambda) N_0(\lambda_{pump}) = \sigma_{abs,0} L_{30}(\lambda) N_0(\lambda_{pump}), \\
g(\lambda, \lambda_{pump}) &= \sigma_{em}(\lambda) N_2(\lambda_{pump}) = \sigma_{em,0} L_{21}(\lambda) N_2(\lambda_{pump}),
\end{aligned} \quad (3)$$

are related to the total refractive index of the gain-composite $n'$ through

$$\frac{\alpha - g}{2} = \frac{\omega}{c} n'', \qquad n'' = \frac{\varepsilon_r''}{2n'} \quad (4)$$

Thus, by Eqs. (1)-(4), the imaginary part of the permittivity of the gain-composite is given by

$$\varepsilon_r''(\lambda) = \frac{n'\lambda}{2\pi} \frac{N}{1 + L_{abs}(\lambda_{pump})|\bar{E}_{pump}|^2} \begin{bmatrix} \sigma_{abs,0} L_{abs}(\lambda) \\ -\sigma_{em,0} L_{em}(\lambda) L_{abs}(\lambda_{pump})|\bar{E}_{pump}|^2 \end{bmatrix}. \quad (5)$$

For realistic parameters (see below), $\varepsilon_r'' = O(10^{-2})$, and typically much smaller. The contribution of the emitters to the *real* part of the permittivity of the composite gain material is of the same order. However, compared with the permittivity of the host, it is much smaller and thus, non-negligible only for very high emitter concentrations [35]. In those cases, this contribution can be computed through Kramers-Krönig relations [18,34] or measured experimentally [35]. In what follows, we ignore this contribution and set $n' = n_{host}$.

The pump-probe configuration requires a two-step solution based on the permittivity (Eq. (5)). The first stage of the solution process is to solve the nonlinear equations governing the pump distribution (at $\lambda = \lambda_{pump}$, preferably equal to $\lambda_{30}$). Previous works assumed a certain pump-field distribution rather than solving the actual nonlinear problem. This approximation may be justified, e.g., for uniform pumping, or for pumping fields much stronger than the saturation field. However, in many cases, in particular, for optical pumping (which was employed in all experimental studies so far), the pumping intensity is limited by the damage threshold of the structures. Thus, complete saturation may be difficult to attain. Consequently, the population, and hence the gain and permittivity, may have a certain degree of non-uniformity. The second stage of the solution process involves solving the equations governing the probe light distribution. Following the discussion above, for a sufficiently weak probe light, those equations are linear, yet, the permittivity may be space-dependent through the parametric dependence on $E_{pump}$. In this paper, we solve the nonlinear problem exactly and discuss the effect of a spatial distribution of the gain on the loss compensation.

## 3. Simulations

The simulations were performed using the frequency-domain solver of the commercial package COMSOL Multiphysics. The NIM design studied here is the fishnet structure, a sandwich structure of two perforated thin metal films separated by a dielectric spacer layer (see e.g., [9,10] and references therein). In the current design (see Fig. 1(c)), the fishnet structure consists of two 50-nm Ag layers separated by a 40-nm dielectric spacer and a thick glass substrate. The sample is coated with a 40-nm layer of gain-composite; this composite also fills the perforations of the fishnet. The unit cell period is 280 nm and the perforation has the typical stadium-type shape with dimensions of 50 nm by 70 nm. The incident light is polarized along the longer side of the perforation. The Ag permittivity is modeled with a causal Drude-Lorentz model [54], and the index of refraction of the dielectric and host materials are chosen to be $1.62$.

We assume the conservative values of $N = 1.2 \times 10^{19} \, \text{cm}^{-3} = 0.012 \, \text{nm}^{-3}$, $\sigma_{abs,0} = 2.7 \times 10^{-16} \, \text{cm}^2 = 0.027 \, \text{nm}^2$, and $\sigma_{em,0} = 0.6 \times 10^{-16} \, \text{cm}^2$, in the numerical simulations below, where we also choose $\lambda_{30} = 718 \, \text{nm}$, $\lambda_{21} = 746 \, \text{nm}$. The absorption and emission cross-sections are assumed to have a Lorentzian shape with an inverse width of 15 fs. These values give rise to a maximal gain coefficient of $g \approx 720 \, \text{cm}^{-1}$. All the above parameters are characteristic of dye molecules [22,34,36–38,42,51]. We choose $\left|E_{pump,inc} / E_{sat}\right| = 5$, which should be below the damage threshold but yet sufficiently higher than $E_{sat}$ in order to obtain sufficiently high gain.

We study three different samples in this work. Sample 1 is a purely passive device, which serves as a reference for the loss compensation. In sample 2, gain emitters are incorporated in the coating layer and the perforations (as shown in Fig. 1(c)) and in sample 3, emitters are also incorporated into the dielectric spacer layer.

### 3.1 Pump simulations

Figure 2 shows the pump-field distribution in two different cross-sections for sample 3 [55]. The pump-field in the first cross-section, from the middle of the spacer layer, is extremely non-uniform and varies between $E_{pump} \approx 0$ and $E_{pump} \approx 15 E_{sat} = 3 E_{pump,inc}$. However, the population of the lasing level is fairly uniform, being close to saturation ($N_0 \ll N_2 \simeq N$) over most of the cross-sectional area. Indeed, this occurs because the incident field is already five times higher than the saturation field; furthermore, the pump-field is locally enhanced with respect to the incident field $E_{pump,inc}$ by up to a factor of three (Fig. 2(a)). Nevertheless, we note that since the pump-field drops below one at the edges of the (quarter) unit cell, those regimes are weakly pumped. Therefore, the pumping in this regime does not contribute to the gain, but rather only serves to reduce the absorption.

The pump-field in the second cross-section, which is within the coating layer above the fishnet structure, is more uniform but significantly less enhanced. Consequently, although the pump-field levels are well above the saturation everywhere in this cross-section, the gain is quite non-uniform.

Overall, for the pumping levels discussed above, the effect of gain non-uniformity amounts to a change of a few percents in the far-field spectra and the effective parameters (see below). Further increase of the pumping level to $|E_{pump,inc}/E_{sat}| = O(10)$, shows that the gain becomes completely uniform at this level. This result is not unexpected; however, such pumping levels may be close to the damage threshold of the metals and thus could be impractical for realistic applications.

*3.2 Probe simulations*

The solution of the pump-field shown above is now substituted into Eq. (5) in order to determine the space-dependent permittivity experienced by the probe light. Figure 3(a) shows the transmission $T$, reflection $R$ and absorption $A$ coefficients obtained for probe light incident on the three samples. For sample 1, a clear resonance dip is seen in the reflection/absorption spectra at $\lambda \approx 745\,\text{nm}$. The emitters in samples 2 and 3, chosen to operate at approximately that same wavelength ($\lambda_{21} = 746\,\text{nm}$), give rise to a reduction of the absorption and to simultaneous reduced reflection and increased transmission. In particular, in Fig. 3(b) we show that the absorption losses are reduced by, at most, ~4% for sample 2. However, the losses are reduced by much higher values, up to 20%, for sample 3 in which the emitters occupy the spacer layer as well as the overcoat layer.

Further increases in the gain give rise to complete and even over-compensation (A < 0, data not shown), hence opening the way to transparency and even to lasing.

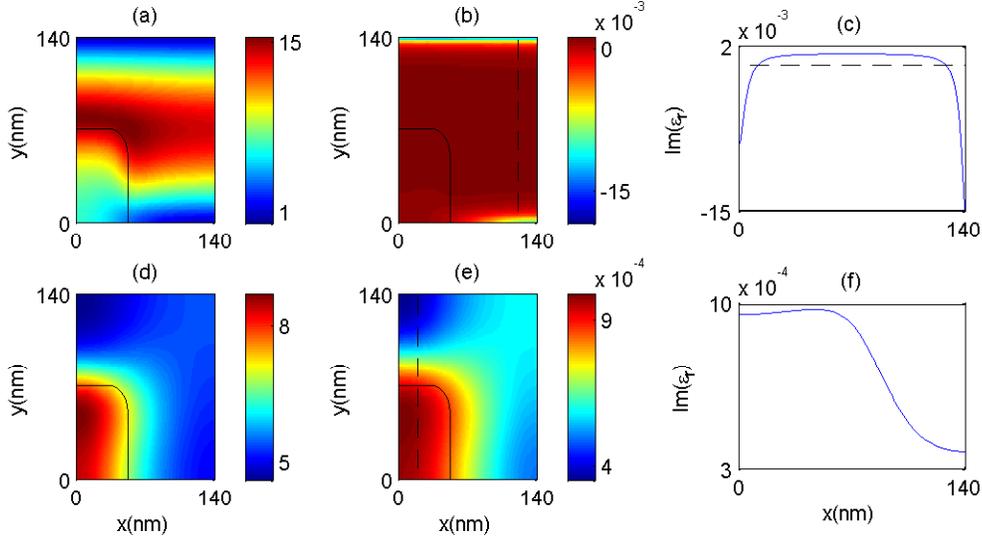

Fig. 2. (Color online) (a) Pump-field normalized by the saturation field $|E_{pump}/E_{sat}|$ and (b) imaginary part of the relative permittivity $\varepsilon_r^"$ in a quarter unit cell for sample C for a cross-section in the middle of the spacer layer. The perforation is outlined by the solid line. (c) The permittivity in a cross-section along the dashed line in (b). The dashed line represents the borderline between absorption and gain ($\varepsilon_r^" = 0$). (d)-(f) Same as (a)-(c), respectively, for a cross-section 15 nm above the upper metal layer. Here, $\lambda_{pump} = \lambda_{30} = 718\,\text{nm}$.

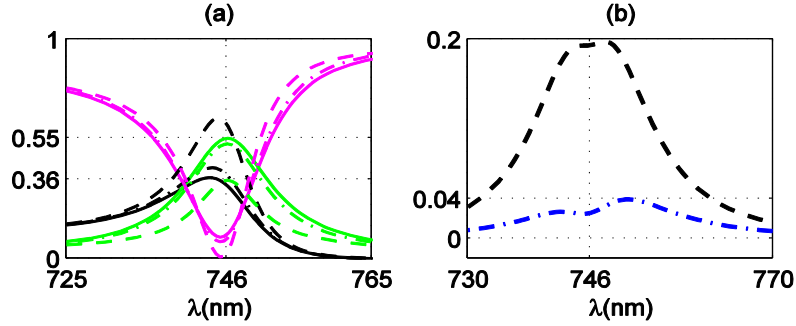

Fig. 3. (Color online) (a) Transmission (black), reflection (magenta) and absorption (green) for a fishnet structure without dye (solid line), with pumped dye molecules in a coating layer (dash-dotted line) and with pumped emitters in coating and spacer layers (dashed line). (b) Reduction in absorption for sample 2 (dash-dotted blue line) and sample 3 (dashed black line).

*3.3 Retrieval of effective parameters*

We turn our attention to the effect of the gain on the effective parameters of the fishnet structure [56], focusing on sample 3 only. Figure 4(a) shows that the real part of the index of refraction attains a 25-nm-wide band of $n'_{eff} \approx -1.67$ centered at $\lambda = 760\,\text{nm}$. Away from that regime, $n'_{eff}$ becomes less negative and approaches $n'_{eff} \approx 0$. The gain leads to a small reduction of $n'_{eff}$ around the resonance, as well as to a reduction of the extinction coefficient $n''_{eff}$ for $\lambda < 760\,\text{nm}$. Accordingly, Fig. 4(b) shows that in the regime $\lambda < 760\,\text{nm}$ there is an improvement of the figure of merit (FOM), defined as $-n'_{eff}/n''_{eff}$, through *both* parameters. The maximal FOM, improved from $\approx 3$ to $\approx 9$, is attained at $\lambda \approx 748\,\text{nm}$ where $n''_{eff}$ approaches zero. Further increases in the gain may yield additional increases in the FOM toward infinity (for $n''_{eff} = 0$) and even amplification ($n''_{eff} < 0$). We note, however, that further away from the resonance, both the index of refraction and the extinction coefficient increase due to the gain, thus corresponding to a lowered FOM. These specific simulations are an example showing that a low-loss NIM regime is feasible, and that it can be isolated in wavelength from adjacent regimes of high losses, in agreement with the predictions of [57].

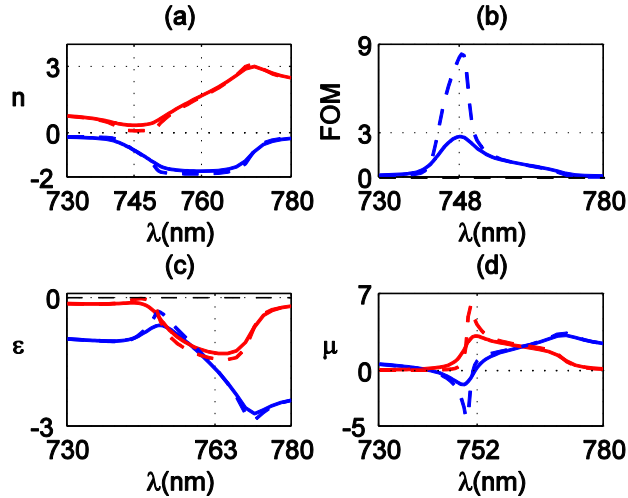

Fig. 4. (Color online) Real parts (blue) and imaginary parts (red) of retrieved effective parameters for samples 1 (solid lines) and 3 (dashed lines).

Calculation of the effective permittivity $\varepsilon_{eff}$ and permeability $\mu_{eff}$ show that the magnetic and electric resonances of the structure under investigation are close to each other, resulting in a complicated lineshape of the combined resonant and anti-resonant responses [58]. Indeed, the effective permittivity exhibits an anti-resonant shape at $\lambda \approx 763$ nm, and the effective permeability exhibits a complicated combination of the resonant and anti-resonant response, as shown in Figs. 4(c) and 4(d), respectively. Both the electric and magnetic plasmonic resonances undergo narrowing and sharpening due to the gain, as expected.

## 4. Analysis of the effective gain

The levels of compensation and the large difference between the compensation in samples 2 and 3 are higher than what can be expected from a simple estimate based on the bulk gain coefficient and the volume occupied by the emitters. Therefore, in order to explain the effect of the gain on the NIM performance and the source of the difference between the performance of samples 2 and 3, we study the absorbed and generated power density at the various structural components of the fishnet. The power absorbed/generated by the Ag/gain-composite is given by

$$Q_{Ag/em} = \frac{1}{2}\omega \int \varepsilon_0 \varepsilon_r''(\omega, \vec{x}) |E(\omega, \vec{x})|^2 d^3x, \qquad (6)$$

where the integration is performed over the corresponding regions occupied by the Ag\gain-composite. The contribution of the power from Eq. (6) to the absorption coefficient $A$ will then be given by

$$A_{st} \equiv \frac{I_{Ag/em}}{I_{inc}} = \frac{Z_0 \omega}{S} \int \varepsilon_0 \varepsilon_r''(\omega, \vec{x}) \left|\frac{E(\omega, \vec{x})}{E_{inc}}\right|^2 d^3x, \qquad (7)$$

where $I_{Ag/em} = Q_{Ag/em}/S$ is the intensity absorbed/emitted in the Ag\gain-composite and $I_{inc} = |E_{inc}|^2/2Z_0$ is the incident intensity, with $S$ being the fishnet (quarter) unit cell area and $Z_0$ the vacuum impedance.

In fact, we show now that the rate equations [51,52] can also be used to derive $A_{em}$. Indeed, recall that the rate of photon *density* emitted through stimulated emission is given by $\sigma_{em} N_2 I / \hbar \omega$. Accordingly, the rate of the *total* number of photons emitted through stimulated emission from the spatial domains occupied by the gain medium is

$$\dot{n}_{st} = \frac{n_{host}}{2Z_0} \frac{\sigma_{em}(\omega) \int N_2(\vec{x}) |E(\omega, \vec{x})|^2 d^3x}{\hbar \omega}.$$

The rate of photons incident on a fishnet (quarter) unit cell is $\dot{n}_{inc} = SI_{inc}/\hbar \omega$. Then, by Eqs. (3)-(4),

$$A_{st} \equiv \frac{\dot{n}_{st}}{\dot{n}_{inc}} = n_{host} \frac{\sigma_{em}(\omega) \int N_2(\vec{x}) |E(\omega, \vec{x})|^2 d^3x}{S|E_{inc}|^2} = \frac{Z_0 \omega}{S} \int \varepsilon_0 \varepsilon_r''(\omega, \vec{x}) \left|\frac{E(\omega, \vec{x})}{E_{inc}}\right|^2 d^3x,$$

which is identical to Eq. (7).

Equation (7) shows the crucial role played by the enhanced local-field in the regions occupied by the emitters; thus, it explains the generally efficient compensation and, in particular, the improved loss-compensation when emitters are placed inside the spacer layer (sample 3 vs. sample 2, Fig. 3(b)). Indeed, the enhancement is much stronger inside the spacer layer than in the perforation and the coating (see e.g., Figs. 5(a) and 5(b), shown for sample 3 at $\lambda = 746$ nm [55]). Since the difference between the peak enhancements in the spacer and coating is ~6, an emitter placed in the spacer layer may be up to 36 times more effective than an emitter placed in the coating. More generally, Fig. 6(a) shows that the local-fields averaged over the total region occupied by the emitters are enhanced ($\rho_2 \equiv \sqrt{\int |E(\vec{x},\lambda)|^2 d^3x / \int |E_{inc}|^2 d^3x} > 1$) with respect to the incident powers across the whole spectrum of interest.

Specifically, Eq. (7) shows that optimal loss-compensation requires an overlap of the gain with the strong local-fields in both space and frequency. First, this implies that the spectral regime in which efficient compensation can be achieved is limited not only by the spectral width of the gain, but also by the spectral regime in which significant near-field enhancement is achieved. In general, as in the current case, the gain spectrum (or equivalently, $A_{st}(\lambda) = A_{em}(\lambda)$) is wider than the spectral width of the strongly enhanced fields, see Fig. 6(b). Therefore, the response of the system to the gain is dominated by the near-field enhancement and is less sensitive to the exact location of emitter resonance.

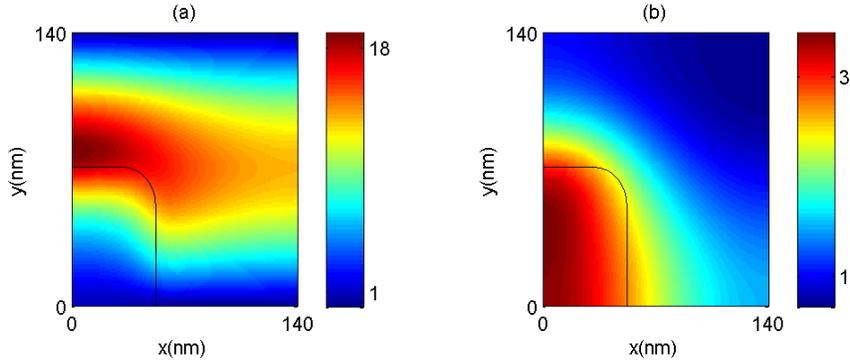

Fig. 5. (Color online) (a) E-field distribution in a quarter unit cell for sample 3 for a cross-section located in the middle of the spacer layer. (b) Same as (a) for a cross-section 15 nm above the fishnet. Here, $\lambda = 746$ nm.

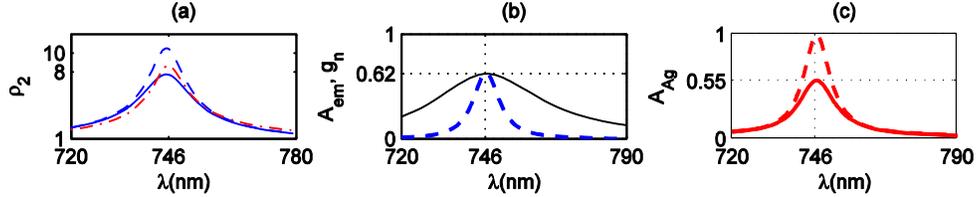

Fig. 6. (Color online) (a) Average E-field enhancement $\rho_2 \equiv \sqrt{\int |E(\vec{x},\lambda)|^2 d^3x / \int |E_{inc}|^2 d^3x}$ for sample 1 (solid blue line) and sample 3 (dashed blue line). Also shown is the maximal enhancement normalized by a factor of 10 (red dash-dotted line). (b) Generated power density in the regions occupied by the gain-composite (dashed, sample 3) compared with the normalized average gain profile $\left(g_n = \max_\lambda (A_{Ag})(g / \max_\lambda (g))\right)$ (black solid line). (c) Absorbed power density in the Ag fishnet layers without dye (solid, sample 1) and with dye (dashed, sample 3).

Second, Eq. (7) shows that when the gain is fairly uniform (i.e., for sufficiently strong pumping, see e.g., Fig. 2), the total compensation is given by $A_{st} \sim \int |E(\lambda,\vec{x})|^2 d^3x$, i.e., it is simply proportional to the average enhancement. Thus, the correct estimate of the loss-compensation should be based on the *average* field enhancement rather than on the *maximal local*-field enhancement. Those enhancements may differ both in magnitude and in spectral profile. Indeed, in our example, the maximal local enhancement is about 10 times higher than the average enhancement, see Fig. 6(a); thus, it may lead to an overestimate of the effective gain. As noted in previous studies [20,46,47], the enhancement can be regarded as an effective increase of the gain coefficient with respect to its value measured in bulk media. In our example, the average enhancement amounts to an increase of the gain coefficient $g$ by a factor of up to $|E/E_{inc}|^2 \approx 10^2$ at $\lambda \approx 746\,\text{nm}$, so that $g_{eff} \approx 100g \approx 72{,}000\,\text{cm}^{-1}$. This effect is the counterpart of the enhanced absorption of plasmonic devices, which, as mentioned in the introduction, attracts so much scientific interest. It should not be confused, however, with enhanced fluorescence due to the Purcell effect (see Section 5).

Third, we note that the wavelength of maximal enhancement (and thus, the peak of $A_{em}$) is dictated by the exact lineshapes of the effective permittivity and permeability, and does not necessarily coincide precisely with the wavelength at which the refractive index is most negative. Indeed, in the example discussed here, maximal enhancement is attained at $\lambda \approx 746\,\text{nm}$ (Fig. 6(b)), while the most negative index is attained at $\lambda \approx 760\,\text{nm}$ (Fig. 4(a)). For that reason, the maximal improvement of the FOM is attained at $\lambda \approx 748\,\text{nm}$, close to the wavelength of maximal average enhancement.

Finally, it should be noted that an estimate of the required gain based on Eq. (7) overlooks the effect of the emitters on the losses in the metal. Indeed, the presence of the emitters gives rise to an overall enhancement of the near-fields, and in particular, in the metal. Consequently, the absorption losses increase with respect to the emitter-free case (sample 1). For example, in Fig. 7, we compare the electric field distribution for samples 1 and 3 in a cross-section located in the middle of the upper metal layer. One can see that the presence of the gain does not alter the penetration of the fields into the metal nor cause any significant redistribution of the fields. Rather, it gives rise to an overall increase of the field magnitude. Consequently, the compensation offered by the emitters amounts to ~62% of the incident light intensity (Fig. 6(b)). This means that more photons are produced by stimulated emission from the emitters than those absorbed in the metal in the absence of the emitters (~55% of the incident light, see Fig. 4(a) or 6(c)). However, the structure with the emitters (sample 3) is still absorptive (A ~36% or $n''_{eff} > 0$, see Fig. 3(a) and 4(a), respectively) because the losses in the metal increase to more than 90% of the incident light due to the presence of the emitters (Fig. 6(c)).

## 5. Discussion and outlook

We have presented frequency-domain simulations of light propagation in plasmonic nanostructures with embedded nonlinearly saturable gain material under optical pumping.

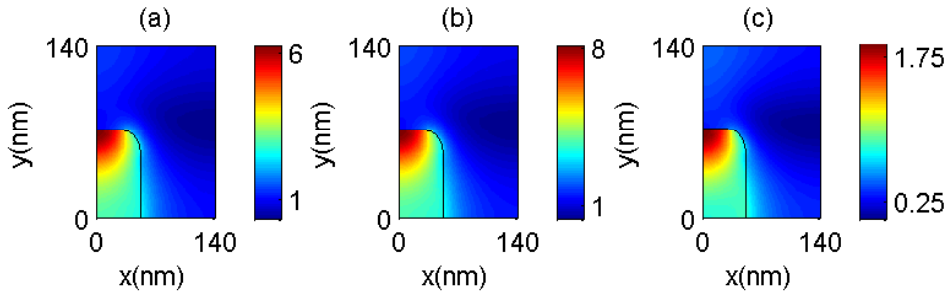

Fig. 7. (Color online) (a) Same as Fig. 5(a) for a cross-section in the middle of the upper Ag layer for sample A. (b) Same as (a) for sample C. (c) Difference of the field maps of (a) and (b).

To the best of our knowledge, no such simulations were performed before in the context of plasmonic nanostructures. We focused on the case of a weak probe beam (small-signal gain), in which the pump equations involve a saturable nonlinear absorption and the equations governing the probe light are strictly linear; this is the most efficient and relevant case for applications. One may argue that near the resonance, the probe-field experiences strong local enhancement, see e.g., Fig. 5(a), hence it may be comparable in magnitude to the pump-field. Nevertheless, since the pump-field is itself enhanced, even if to a lesser extent (see e.g., Fig. 2(a)), the condition for neglecting the probe effect on the population, $q_i |E|^2 \ll |E_{pump}|^2$, may still hold. Whenever it does not (see e.g., [46,47]), one should take into account the probe effect by using Eq. (1) rather than its approximation. This generalization is straightforward within our formalism.

The solution of the nonlinear equations of the pump-field allowed us to study the non-uniformity of the gain in the regions occupied by the gain-composite. We showed that for a weak probe, although the electric field may be extremely non-uniform, for sufficiently strong fields, the gain system saturates, the non-uniformity diminishes and the gain approaches its maximal value everywhere in the device. However, several effects which have been neglected in the current study may give rise to a stronger non-uniformity. First, as mentioned above, in regimes where the probe is significantly enhanced, the gain may be locally reduced due to local population depletion. Second, note that for simplicity we assumed that the quantum efficiency $q_i$ (or emission cross-section $\sigma_{em}$) and the emitter lifetime $\tau$ are uniform in space, while in practice, they may be locally modified (quenched) due to the proximity of the emitter to the metal (Purcell effect). In fact, in fishnet designs, most of the emitters are in proximity to the metal, especially when those are positioned inside the spacer layer. Hence, quenching may play a significant role as a large portion of the emitter energy will be transferred to several non-radiative mechanisms such as lossy surface waves etc [59,60]. As a consequence, the gain non-uniformity would increase and would have to be taken into account. These effects can be incorporated in our formalism simply by replacing the scalar parameters with space-dependent ones. It should be noted, however, that although the lifetime and quantum yield of emitters in complicated geometries such as the fishnet structure *can* be computed [61], this is a formidable theoretical challenge and the exact dependence cannot be reduced to any simplified form. Moreover, these computations are extremely expensive to perform. Accordingly, assuming that the quantum efficiency $q_i$ and the emitter lifetime $\tau$ are uniform in space, equaling some experimentally-measured average values (not necessarily equal to those in the absence of metal), may be a reasonable and even unavoidable assumption. Nevertheless, it may be worthwhile and even necessary to do so for simpler geometries (e.g., for surface plasmons [37,38], simpler NIM designs [44,62,63], composite superlenses [64] or hyper-lenses [13,14]) or for obtaining agreement with experimental measurements. Importantly, once such models are obtained, their implementation in our nonlinear frequency-domain formalism amounts to only a few percent of overhead computation time, even for the 3D structures under consideration. In that respect, our model may be viewed as a first necessary step towards studying these effects.

We have further shown that the loss-compensation is more efficient than that estimated through the bulk parameters of the gain-composite and that optimal compensation can be obtained if the emitters are incorporated in the spacer layer where the field enhancement is maximal. We explained this result by considering the loss/gain at each structural component rather than at the effective parameters of the structure. Indeed, using exact relations derived from the Poynting theorem and rate equations, we showed that the compensation is directly related to the average near-field enhancement. This shows that in order to exploit the full spectral width of the gain offered by the emitters, it would be beneficial to design metamaterials with average near-field enhancements as spectrally wide as possible, e.g., enhancements based on semi-continuous films [62–64]. Furthermore, we showed that an estimate of the required gain for complete compensation should also take into account the increased losses in the metal due to the field enhancement in the metal.

Overall, exploiting the strong near-field enhancement in the spacer layer allows one to have lossless NIMs without the need to supplement the original structure with a thick gain layer, thus keeping the whole unit cell subwavelength in size. This is a clear advantage over most existing loss-compensated plasmonic metamaterial designs, an advantage originating from the large part of the unit cell in which the field is enhanced. Clearly, however, the fabrication challenge involved in inserting the emitters into the spacer layer is formidable.

The model presented in this paper neglected the effect of spontaneous emission from the emitters on the performance of the NIM. This effect can be incorporated into our model by including a random current source in the regimes occupied by the emitters, see e.g [53,65]. Clearly, the noise created by spontaneous emission is undesirable, however, for a sufficiently strong probe such that the rate of stimulated emission between level 2 and 1 is higher than the rate of spontaneous emission, the effect of the latter can be safely neglected.

Finally, it should be noted that quantum effects such as the change in lifetime and quantum yield discussed above, changes in absorption due to modified surface states [35], quantum coupling between the gain and plasmonic systems [66] and spasing [18] may prove to be crucial for the interpretation of experimental measurements. Those effects will be studied elsewhere.

**Acknowledgments**

We would like to thank Prof. Thomas Klar for useful initial discussions. This work was supported in part by ARO-MURI award 50342-PH-MUR and NSF PREM grant DMR 0611430.